# Frequency comb generation in low-loss, low-stress, high-Q deuterated silicon nitride microring resonators in an 8-inch photonics platform

Y. CAO[1,*], G. F. CHEN[2], C. LAU[1], L. Y. M. TOBING[1,3], S. L. H. JANG[3], Y. F. TSANG[3], J. O. YOO[1,3], Y. T. TOH[3], J. S. GOH[1], L. W. LIM[1,3], C. W. WONG[1,4], D. K. T. NG[1,3], D. T. H. TAN[1,2,3], X. LUO[1,3]

[1]*National Semiconductor Translation and Innovation Centre (NSTIC), 1 Fusionopolis Way, #19-10 Connexis North, Singapore 138632*
[2]*Photonics Devices and System Group, Singapore University of Technology and Design, 8 Somapah Rd, Singapore 487372, Singapore*
[3]*Institute of Microelectronics, Agency for Science, Technology and Research (IME, A*STAR), 2 Fusionopolis Way, #08-02 Innovis Tower,138634 Republic of Singapore*
[4]*Fang Lu Mesoscopic Optics and Quantum Electronics Laboratory, University of California, Los Angeles, CA, USA*
\* *cao_yanmei@.a-star.edu.sg*

**Abstract:** Systematic studies on different SiN films in terms of propagation losses are presented, and deuterated SiN emerges as a good candidate for ultralow loss (< 0.1 dB/cm) and reliability by simple 8-inch process with low thermal budget. Frequency comb generation in high-Q (~1 million) deuterated silicon nitride microring is demonstrated and used for intensity modulated direct detection transmission. Negligible power penalty for 25.78 GBaud/s NRZ and PAM4 is achieved at error rates <$10^{-6}$, below the FEC limit.

## 1. Introduction

Consisting of a series of discrete, equally spaced and phase-locked frequency lines, optical frequency combs have found significant applications in a variety of fields, including atomic clocks [1–3], astronomy [4–6], optical communications [7–10,], ranging [11–14], spectroscopy [15,16], and quantum optics [17,18]. Frequency comb has been studied on various material platforms [19–21], among which silicon nitride (SiN) has emerged as a popular one, owing to its high Kerr nonlinearity, low loss, temperature insensitivity, and low dispersion. Besides, SiN platform is extensively used in microelectronics and photonics fabrication processes, making it an ideal candidate for monolithic integration with active photonic components and high-frequency electronics—key elements for the practical deployment of optical comb sources.

In recent years, the need for ultra-low propagation losses has intensified, in part for their importance for on-chip frequency comb generation and other quantum-related applications [19]. Fabrication-related non-idealities, such as sidewall roughness, material absorption and defects within the film, contribute to insertion losses. The overarching cause for the optical loss in SiN is the absorption originated from the overtone of Si-H bonds due to the use of Silane-based precursor gases during chemical vapor deposition processes. While the hydrogen content can be reduced (but not completely eliminated) with Dichlorosilane ($SiCl_2H_2$) in low pressure chemical vapor deposition (LPCVD), its much higher deposition temperature (>700 °C) introduces high stress that results in film cracking at thicknesses above 400 nm [22,23]. To achieve thicker films, the substrate is typically pre-patterned to minimize the stress buildup [24]. Furthermore, post-annealing steps are often required to purge out hydrogen content from the as-grown SiN film, carried out at 1100–1200 °C temperature for 12 hours due to the slow diffusion of hydrogen [25]. The above requirements for ultralow loss results in a complex fabrication flow which is often not cost and time effective.

In this work, we conduct a systematic study of SiN materials realized by conventional LPCVD at 780 °C and deuterated silicon nitride (SiN:D) deposited by inductively coupled plasma chemical vapor deposition (ICP-CVD) at 250 °C. Using SiN:D films, we fabricate waveguides and micro-ring resonators on a 200 mm wafer. We achieve ultra-low loss waveguides (<0.1 dB/cm) and high *Q*-factor (loaded *Q* of around 1 million) SiN:D micro-ring resonators (MRRs). SiN:D micro-ring resonators are successfully demonstrated for on-chip frequency comb generation, where both C- and L-band pumping are possible. The generated comb spectrum is used for high-speed intensity modulated direct detection (IMDD) transmission over up to 10 km of optical fiber, where error-free IMDD transmission and negligible penalty of 25.78 GBaud/s NRZ and PAM4 are obtained.

## 2. Material development for SiN films

To develop the material platform, SiN films are grown through both the standard LPCVD process usingDichlorosilane ($SiCl_2H_2$) as the precursor and the ICP-CVD process using deuterated silane ($SiD_4$) as the precursor, using the same 200 mm fabrication process flow. The device fabrication starts with the SiN film deposition on a 5 µm thick thermal oxide on silicon substrate, followed by deep ultra-violet (DUV) lithography, dry etching, and the upper cladding deposition of 3 µm $SiO_2$ via plasma enhanced chemical vapor deposition (PECVD) process. Figures 1(a) and 1(b) show the scanning electron micrographs for SiN:D and LPCVD SiN waveguides that exhibit similar sidewall roughness with an average line edge roughness (LER) of 2.3 nm and 2.4 nm, respectively.

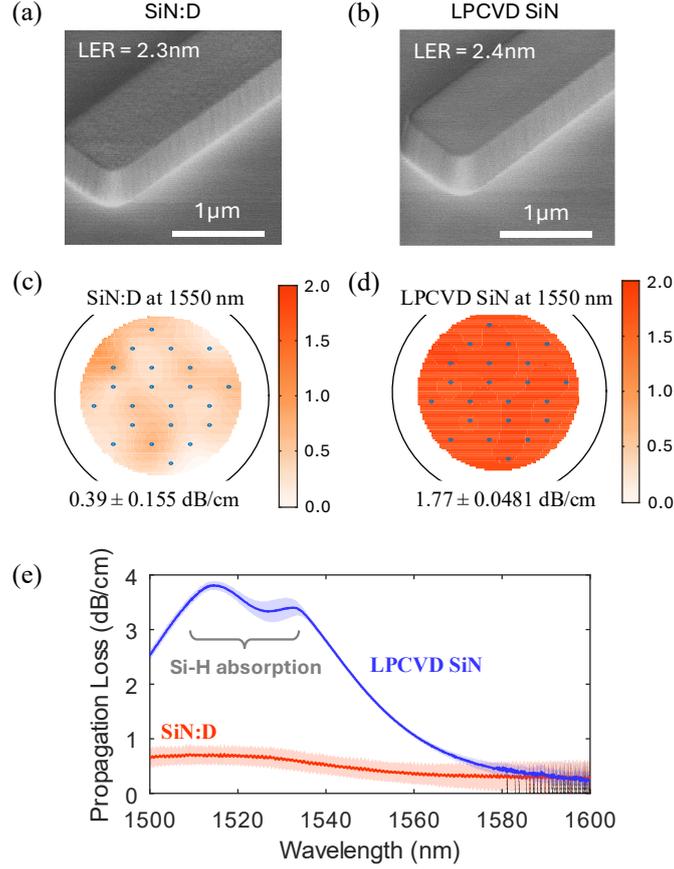

**Fig. 1.** Scanning electron micrographs (SEMs) for the fabricated SiN:D waveguides (a) and LPCVD SiN waveguides (b). Measured wafer-scale propagation loss for SiN:D waveguides (c) and LPCVD SiN waveguides (d) at 1550 nm. (e) Measured propagation loss for SiN:D and LPCVD SiN waveguides across C-band, where the solid lines show the average loss and the shaded areas show the standard deviation of 22 dies across the whole wafer.

SiN waveguides with a thickness of 400 nm and width of 1 µm are fabricated, and wafer-scale measurements in C-band are performed to extract the propagation loss. The measured propagation loss at 1550 nm for SiN:D and LPCVD SiN waveguides are plotted in Fig. 1(c) and 1(d). At 1550 nm, LPCVD SiN waveguides have a higher loss of (1.77 ± 0.0481) dB/cm than SiN:D waveguides loss of (0.39 ± 0.155) dB/cm. The measured waveguide loss across C-band (1500–1600 nm) are shown in Fig. 1(e), in which we observe an absorption dip close to 1520 nm in LPCVD SiN waveguides caused by Si-H bond absorption due to the use of Dichlorosilane.

This mechanism is verified by comparing the loss in O-band using 800 nm wide single-mode waveguides, where all the waveguide loss comes from scattering loss on the sidewall and within the film (caused by voids, clusters, and film granularity). Since the scattering loss on the sidewall are similar in LPCVD SiN and SiN:D waveguides given the same waveguide width and similar LER, the loss difference in O-band is caused by the scattering loss within the film. The measured propagation loss at 1310 nm for SiN:D and LPCVD SiN waveguides are plotted in Fig. 2(a) and 2(b), with a value of (0.25 ± 0.0294) dB/cm in SiN:D waveguides and (0.33 ± 0.0629) dB/cm in LPCVD SiN waveguides. Figure 2(c) shows the waveguide loss in O-band for SiN:D and LPCVD SiN waveguides, in which the two loss spectra are almost overlapping, signifying a similarly low number of voids in both the SiN:D and LPCVD SiN film.

Based on the above comparison, SiN:D is indeed a promising approach for the realization of SiN photonics devices. Incorporation of $SiD_4$ leads to a simplified process flow that enables the realization of ultralow loss SiN without the need for a prolonged high temperature annealing. Owing to its low film stress, the SiN:D film can be much thicker than its LPCVD counterparts, which holds importance to satisfy the anomalous dispersion condition required to generate frequency combs. Despite the higher cost in $SiD_4$ precursor gas, the fabrication flow consists of only a single, straightforward deposition step. Thus, the low thermal budget, low film stress, and much simpler fabrication steps of SiN:D make an attractive choice over LPCVD-based SiN which requires complicated deposition and annealing process.

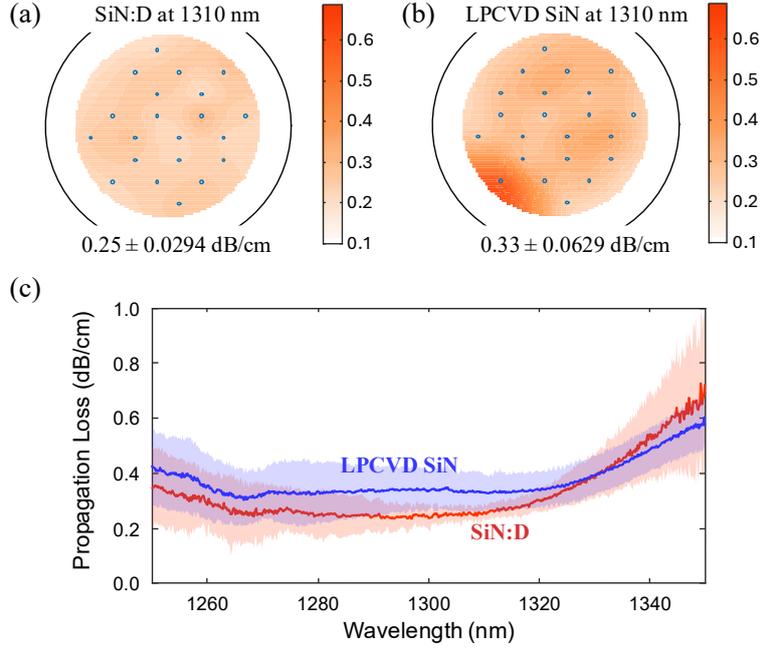

**Fig. 2.** Measured wafer-scale propagation loss for SiN:D waveguides (a) and LPCVD SiN waveguides (b) at 1310 nm. (c) Measured propagation loss for SiN:D and LPCVD SiN waveguides across C-band, where the solid lines show the average loss and the shaded areas show the standard deviation of 22 dies across the whole wafer.

## 3. On-chip frequency comb generation in SiN:D microring resonators

### 3.1 Device design and fabrication

To demonstrate on-chip frequency comb generation, we fabricated microring resonators using the same process flow as discussed in section 2. Specifically, the fabrication starts with an 8-inch silicon wafer with a 5 µm thick thermally grown silicon dioxide ($SiO_2$). SiN:D film with a thickness of 800 nm is deposited via ICP-CVD process using $SiD_4$ precursors at 250 °C, in a single deposition step. Subsequently, the microring resonator (MRR) structure is patterned usings a DUV stepper and transferred to the SiN:D layer via ICP etching. PECVD $SiO_2$ of 3 µm thickness is deposited as the top cladding, also at <400°C temperature.

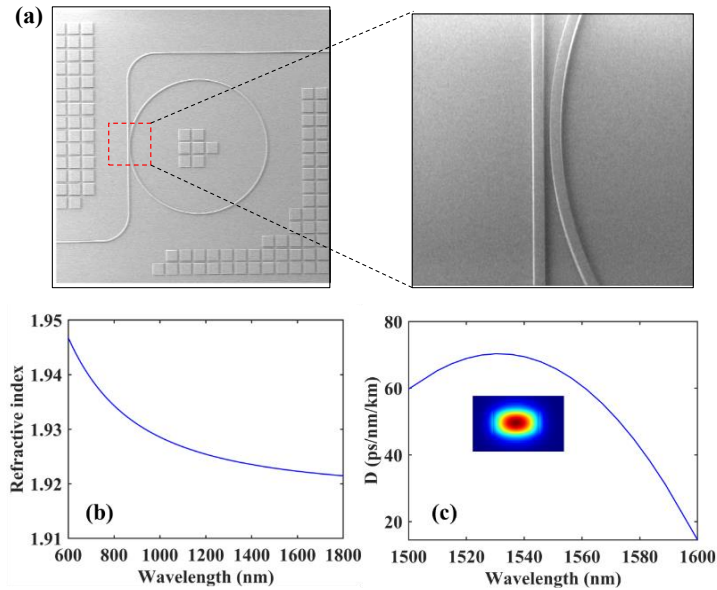

**Fig. 3.** (a) Top-view SEM images of the fabricated SiN:D MRR. (b) Measured refractive index of the SiN:D film by an ellipsometry. (c) Simulated dispersion distribution for the $TE_{00}$ mode in a SiN:D waveguide with width of 2 µm and thickness of 800 nm. The inset shows the simulated $TE_{00}$ mode profile at 1550 nm.

Scanning electron micrographs (SEMs) of the fabricated SiN:D MRR are shown in Fig. 3(a). The refractive index of the SiN:D film is measured by an ellipsometry and plotted in Fig. 3(b), with a value of ~1.92 at 1550 nm, indicating that the material is slightly nitrogen-rich. Figure 3(c) shows the simulated dispersion for the waveguides with width of 2 µm and thickness of 800 nm. With this waveguide design, we ensure all the designed MRRs have anomalous dispersion which is necessary for the generation of various comb states including the primary comb state and bright solitons.

### 3.2 Device characterization and frequency comb generation

MRRs with different radii are fabricated using the same process flow. To characterize the device quality, we first measured the ring transmission in a MRR with a radius of 240 µm, using a TE-polarized tunable laser before coupling into the resonator. As plotted in Fig. 4(a), the Q-factor at each resonance is extracted through Lorentzian fitting, showing loaded Q-factor ($Q_L$) of ~ 1 million and a free-spectral range (FSR) of ~98GHz (~0.83nm). The propagation loss is then estimated based on $TE_{00}$ mode resonance near 1589.758 nm [Fig. 4(b)]. The loaded Q-factor ($Q_L$) is 1.03 ×$10^6$, and the corresponding intrinsic Q-factor ($Q_{int}$) is 4.04 ×$10^6$, as calculated by $Q_{int} = 2 \times Q_L/(1 - \sqrt{T_{res}})$, with $T_{res}$ as the on-resonance transmission. The extracted propagation loss is estimated from intrinsic Q-factor by $\alpha = 4.34 \times 2\pi n_g/(Q_{int}\lambda_{res})$ ~0.085 $dB/cm$ (<0.1 dB/cm), where $n_g$ is the group index, and $\lambda_{res}$ is the resonance wavelength. The propagation loss obtained here is similar to those reported in other SiN:D devices [26–30]. Figures 4(c) and 4(d) show the transmission and $Q_L$ in 160 µm and 100 µm SiN:D MRRs, where around 1 million $Q_L$ and < 0.1 dB/cm waveguide loss are also observed, with a FSR of 148 GHz and 237 GHz, respectively.

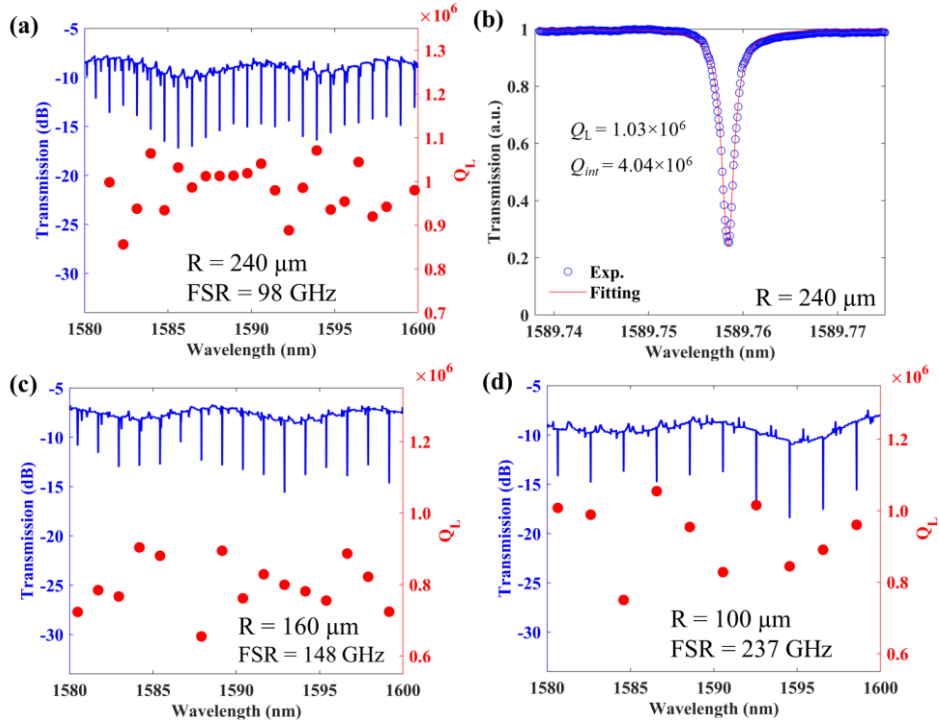

**Fig. 4.** (a) Measured transmission spectrum and the extracted loaded Q-factor, $Q_L$, and (b) the measured and fitted resonance near 1589.758nm for SiN:D MRR with 240 µm in radii. (c) – (d) Measured transmission and $Q_L$ for a SiN:D MRR with a radius of 160 µm and 100 µm.

It has been demonstrated that the threshold power required for on-chip frequency comb generation is inversely proportional to $Q_L^2$ [31]. Thus, it is easier to demonstrate the comb generation using a resonance with a higher $Q_L$. The schematic diagram of the experimental setup used for frequency comb generation is shown in Fig. 5(a). In the measurement, a tunable CW laser is used to launch the pump light, the power of which is amplified by an erbium-doped fiber amplifier (EDFA). Then the pump is tuned to quasi-TE polarization using a polarization controller and edge-coupled into the MRR via a tapered lensed fiber. The fiber-chip coupling loss is estimated to be ~5 dB/facet. The output spectrum of the MRR is split by a power splitter with 10% into a power meter for alignment, and 90% into an optical spectrum analyzer (OSA) for spectrum monitoring. The pump wavelength is tuned manually with a small step (~1 pm) from the blue side toward the red side of the selected resonance, and frequency combs with different states are observed at different pump-resonance detuning positions. The on-chip power used is fixed at 28 dBm.

Figure 5(b) shows the frequency combs generated in a SiN:D MRR with a radius of 160 µm, where the comb spectrum evolves from state I to state V when the pump wavelength is tuned from blue to red side of the selected resonance. From Fig. 5(b), it is observed that the primary combs appear firstly due to four-wave mixing (state I), then sub-combs around the primary combs show up (state II), and more comb lines emerge together to form a gap-free spectrum (state III). Afterwards, the comb spectrum gets denser (state IV) and finally reaches the single soliton state (state V) where the comb spacing equals the FSR of the MRR (~ 148 GHz) and has a smooth $sech^2$ envelope.

The single soliton state is replotted in Fig. 5(c) together with the fitted envelope. The fitting shows a 3-dB bandwidth of 35.1 nm, ranging from 1568.3 to 1603.4 nm. The pump-comb conversion efficiency is calculated to be 4.87% using the same method as stated in Ref. [32]. Frequency combs are also observed in SiN:D MRRs with radii of 100 µm and 240 µm, although only state I to IV is obtained when detuning near some resonances.

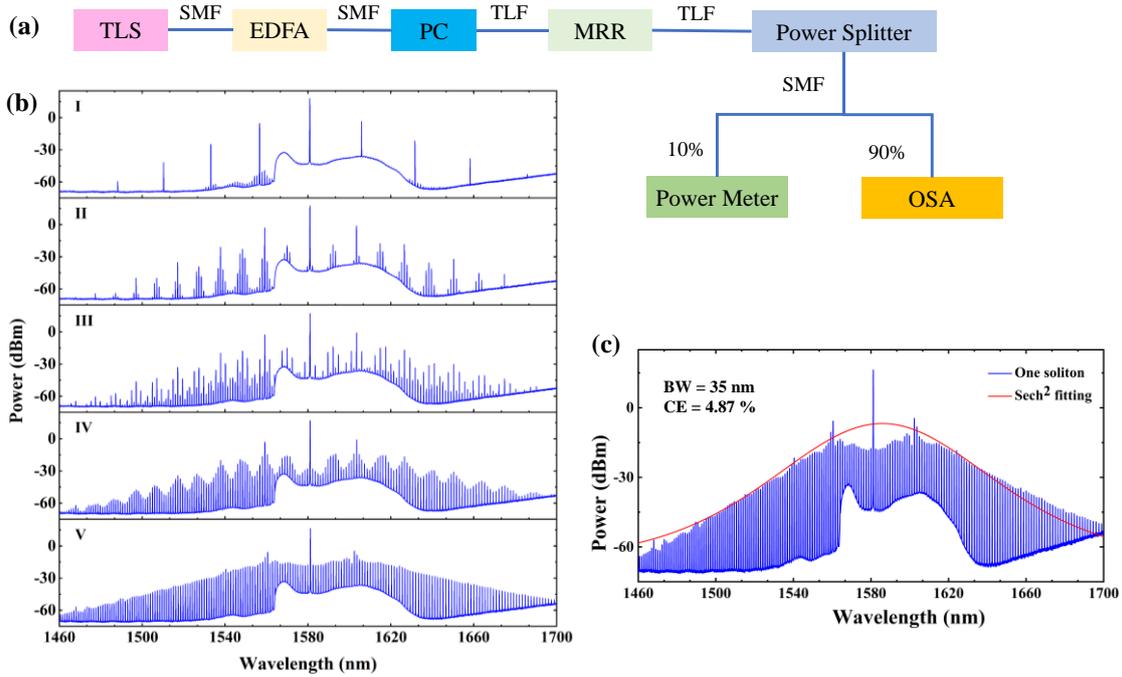

**Fig. 5.** (a) Measured frequency comb generation in a 160 µm SiN:D MRR at different pump-resonance detuning positions. (b) Schematic diagram of the experimental setup for frequency comb generation, where TLS: Tunable Laser Source, EDFA: Erbium-Doped Fiber Amplifier, PC: Polarization Controller, OSA: Optical Spectrum Analyzer, SMF: Single Mode fiber, TLF: Taper Lensed Fiber. (c) Measured comb spectrum (blue line) and its sech$^2$ fitted envelop (red line) of the single soliton state.

**3.3 High-speed transmission using generated frequency comb**

The generated frequency comb in the SiN:D MRR can be applied to high-speed intensity modulated direct detection (IMDD) data transmission. The experimental schematic for IMDD transmission is shown in Fig. 6(a). A data stream with 25.78125 Gbps symbol rate which conforms to the 100 Gigabit Ethernet standard CAUI4 interface with 802.3 Clause 91 RS-FEC is used. As shown in Fig. 6(a), frequency comb is generated in the SiN:D MRR by sweeping an amplified CW laser from the blue side to a preset wavelength in the resonance. The amplification was done using an L-band high power EDFA (EDFA1) to a pump wavelength 1584.79 nm. Various power taps were drawn to wavelength meters, power meters and OSAs as shown for in-experiment optimization purposes. The comb spectrum (primary comb state) measured at point "A" in the schematic in Fig. 6(b). The first sideband comb line, 1565.24 nm is filtered by a band-pass filter (BPF1), amplified by a second EDFA (EDFA2) and bandpass filtered again by a second BPF (BPF2) to remove ASE noise. It is then modulated via a Mach-Zehnder Modulator and fed through a 2-km long Single Mode Fiber (SMF). The 2-km SMF mimics signal propagation in intra-data center communications and is finally back-converted to an electrical signal via a photoreceiver for bit error rate (BER) and eye diagram measurement. Another set of control experiments is repeated via a regular 100 kHz linewidth laser of equivalent power level to the comb line to represent Back-to-Back (B2B) measurements. The B2B characterization uses the laser before BPF1 which is subsequently modulated with the same data stream.

Figure 6(c) shows the BER measurement of comb-modulated and control B2B transmission using NRZ and PAM-4 symbols. At the $10^{-4}$ level, the PAM-4 transmission has negligible power penalty whereas the NRZ has a marginal 0.1 dB penalty compared to the B2B control readings, which suggests that the comb generated from this device may be used efficiently for IMDD transmission. The lowest BER obtained is <$10^{-6}$, which is well below most FEC limits. Eye diagrams obtained from the comb-modulated and control B2B transmission both show open eyes for NRZ data. Eye closure between level 2 and level 3 are observed in both PAM-4 B2B and comb-modulated PAM-4 transmission. This is attributed to amplified noise introduced by EDFA2 which was not completely suppressed by BPF2 which suggests it is not an artifact originating from the Turing comb.

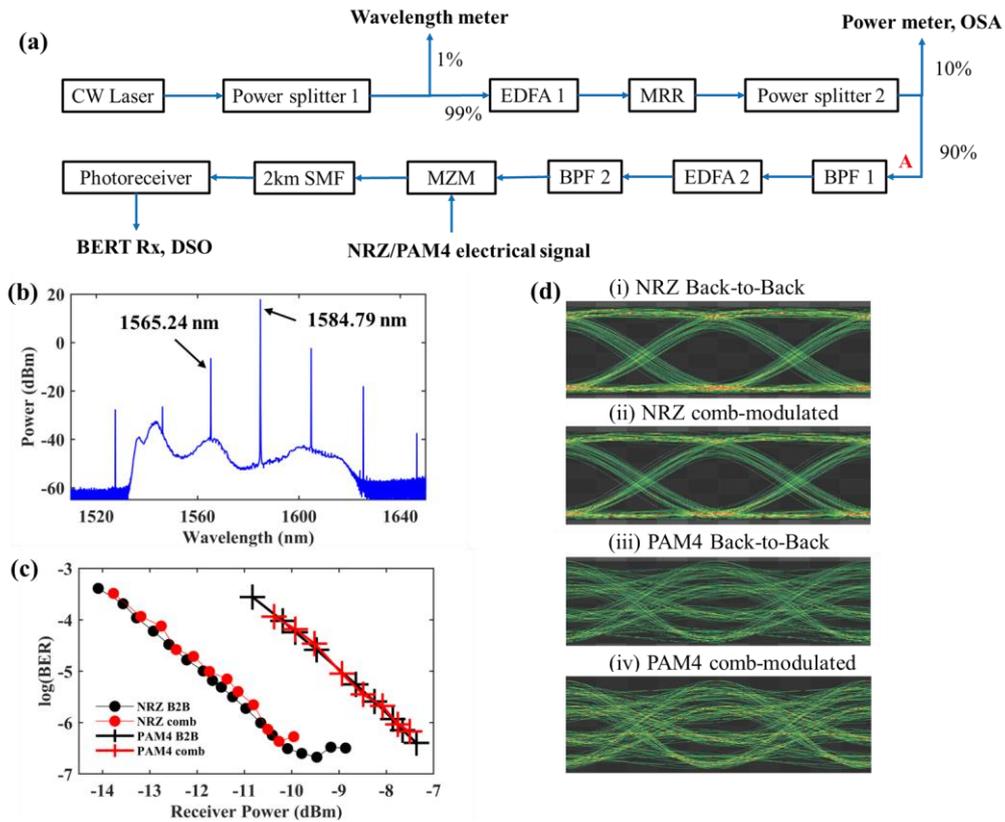

**Fig. 6.** (a) Experimental schematic for IMDD transmission. BPF: Bandpass Filter. MZM: Mach Zehnder Modulator. NRZ: Non-Return-to-Zero. PAM4: Pulse Amplitude Modulation 4levels. BERT Rx: Bit Error Rate Tester Receive port. DSO: Digital Sampling Oscilloscope. (b) Measured comb spectrum in a 160 μm SiN:D MRR at point A. (c) Bit-Error Rate measurement of comb modulated transmission. Red: comb-modulated. Black: back-to-back laser being used. Dots: 25.78125 Gbps NRZ format used. Plus sign: 25.78125 GBaud/s PAM4 format used. (d) Eye diagram readings: (i) NRZ Back-to-back (ii) NRZ Comb-modulated (iii) PAM4 Back-to-back (iv) PAM4 comb-modulated.

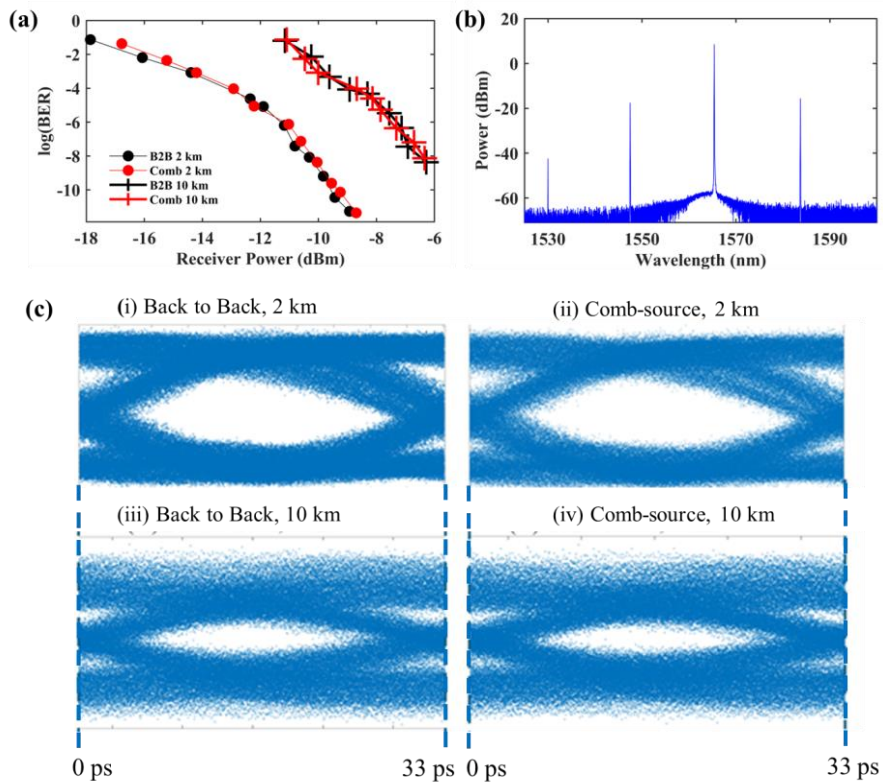

**Fig. 7.** (a) The measured Bit Error Rate for 30Gbps NRZ data over 2km and 10km of optical fiber. (b) The generated primary comb when pumped at 1565nm. The comb line at 1550nm is used for high-speed measurements. (c) The measured eye diagrams for B2B and comb modulated data.

In a similar experiment, frequency comb is generated by pumping the MRR in the C-band (1565 nm). Figure 7(b) shows the measured spectrum of the generated comb. The comb line generated at 1550 nm is used to modulate 30 Gbps NRZ data prior to transmission over optical fibers of 2 km and 10 km in length. Figure 7(a) shows the measured BER of 30 Gbps NRZ data modulated and transmitted over the optical fiber. It may be observed that a similar BER is achieved for both 2 km and 10 km of optical fiber when comparing the comb-modulated and B2B characterization. Figure 7(c) shows the measured eye diagrams using a 30 Gbps NRZ data for the B2B and comb-modulated setups and it may be observed that similar eye diagrams are observed for both. These characterizations showcase that the primary comb state generated in the SiN:D resonators has low noise and is effective as a vessel for the transmission of high-speed data.

## 4. Conclusion

We have systematically studied silicon nitride material deposited via LPCVD and ICP-CVD processes on an 8-inch platform. The device properties are characterized using a wafer-scale characterization tool. It is found that deuterated SiN deposited via ICP-CVD is a highly promising platform owing to its ultra-low losses, reliability, and low thermal budget, enabling a significantly simplified integration process compared to LPCVD SiN. Importantly, low-loss SiN films which are sufficiently thick to exhibit anomalous dispersion at the C- and L-bands may be grown using this technique, of primary importance for a plethora of applications in quantum optics, nonlinear optics and in particular, frequency comb generation.

Frequency comb generation is demonstrated using SiN:D MRRs with high-Q ($Q_{int}$ of $4.04 \times 10^6$) and low loss (loss of 0.085 dB/cm), using a single CW laser as the pump. Especially, single soliton state comb is observed with a 3-dB bandwidth of 35.1 nm, ranging 1568.3 nm to 1603.4 nm, with a pump-comb conversion efficiency of 4.87%. The primary comb state is successfully generated by pumping at both the C-band and L-band. The frequency combs generated in the SiN:D MRR are further applied successfully for high-speed IMDD transmission, where a negligible power penalty for 25.78 GBaud/s NRZ and PAM4 data is obtained at BERs less than $10^{-6}$, well below most FEC limits.

### Acknowledgements

We acknowledge the funding support from the National Semiconductor Translation and Innovation Centre. Dawn T. H. Tan acknowledges funding from the National Semiconductor Translation and Innovation Center (M24W1NS004) and the National Research Foundation Investigatorship.

### Disclosures

The authors declare no conflict of interest.

### REFERENCES

1. T. Fortier and E. Baumann, "20 years of developments in optical frequency comb technology and applications," Commun. Phys. **2**, 153 (2019).
2. Zachary L. Newman, Vincent Maurice, Tara Drake, *et al.*, "Architecture for the photonic integration of an optical atomic clock," Optica **6**(5), 680–685 (2019).
3. A. L. Gaeta, M. Lipson, and T. J. Kippenberg, "Photonic-chip-based frequency combs," Nat. Photon. **13**, 158–169 (2019).
4. E. Obrzud, M. Rainer, A. Harutyunyan, *et al.*, "A microphotonic astrocomb," Nat. Photon. **13**, 31–35 (2019).
5. MG. Suh, X. Yi, YH. Lai, *et al.*, "Searching for exoplanets using a microresonator astrocomb," Nat. Photon. **13**, 25–30 (2019).
6. T. Steinmetz, T. Wilken, C. Araujo-Hauck, *et al.*, "Laser Frequency Combs for Astronomical Observations," Science **321**(5894),1335–1337 (2008).
7. H. Hu and L. K. Oxenløwe, "Chip-based optical frequency combs for high-capacity optical communications," Nanophotonics **10**(5), 1367–1385 (2021).
8. B. Corcoran, M. Tan, X. Xu, *et al.*, "Ultra-dense optical data transmission over standard fibre with a single chip source," Nat. Commun. **11**, 2568 (2020).
9. P. Xing, G. F. R. Chen, H. Gao, *et al.*, "Microresonator Frequency Comb Based High-Speed Transmission of Intensity Modulated Direct Detection Data," Nanophotonics **11**, 3269–3280 (2022).
10. K. Y. K. Ong, A. A. Rahim, X. X. Chia, *et al.*, "High-speed data transmission over a microresonator frequency comb with dispersion compensation for augmented data rates and reach," Nanophotonics **13**, 2367–2378 (2024).
11. M.-G. Suh and K. J. Vahala, "Soliton microcomb range measurement," Science **359** (6378), 884–887 (2018).
12. P. Trocha, M. Karpov, D. Ganin, *et al.*, "Ultrafast optical ranging using microresonator soliton frequency combs," Science **359**(6378), 887–891 (2018).
13. J. Riemensberger, A. Lukashchuk, M. Karpov, *et al.*, "Massively parallel coherent laser ranging using a soliton microcomb," Nature **581**, 164–170 (2020).
14. J. Wang, Z. Lu, W. Wang, *et al.*, "Long-distance ranging with high precision using a soliton microcomb," Photon. Res. **8**(12), 1964–1972 (2020).
15. M.-G. Suh, Q. Yang, K. Yang, *et al.*, "Microresonator soliton dual-comb spectroscopy," Science **354**(6312), 600–603 (2016).
16. N. Picqué and T. W. Hänsch, "Frequency comb spectroscopy," Nat. Photon. **13**, 146–157 (2019).
17. Z. Yang, M. Jahanbozorgi, D. Jeong, *et al.*, "A squeezed quantum microcomb on a chip," Nat. Commun. **12**, 4781 (2021).
18. Y. Zhang, M. Menotti, K. Tan, *et al.*, "Squeezed light from a nanophotonic molecule," Nat. Commun. **12**, 2233 (2021).
19. L. Chang, S. Liu, and J. E. Bowers, "Integrated optical frequency comb technologies," Nat. Photon. **16**, 95–108 (2022).
20. F. Hu, A. K. Vinod, W. Wang, *et al*, "Spatio-temporal breather dynamics in microcomb soliton crystals," Light: Sci. Appl. **13**, 251 (2024).

21. T. Herr, K. Hartinger, J. Riemensberger, *et al*., "Universal formation dynamics and noise of Kerr-frequency combs in microresonators," Nat. Photon. **6**, 480–487 (2012).
22. X. Ji, S. Roberts, M. C. Zanarella, *et al*., "Methods to achieve ultra-high quality factor silicon nitride resonators," APL Photonics **6**(7), 071101 (2021).
23. J. Liu, G. Huang, R.N. Wang, *et al.*, "High-yield, wafer-scale fabrication of ultralow-loss, dispersion-engineered silicon nitride photonic circuits," Nat. Commun. **12**, 2236 (2021).
24. M. H. P. Pfeiffer, C. Herkommer, J. Liu, *et al*., "Photonic damascene process for low-loss, high-confinement silicon nitride waveguides," IEEE J. Sel. Top. Quantum Electron., **24**(4), 1–11 (2018).
25. K. Luke, Y. Okawachi, M. R. E. Lamont, *et al.*, "Broadband mid-infrared frequency comb generation in a $Si_3N_4$ microresonator," Opt. Lett. **40**(21), 4823–4826 (2015).
26. Y. Xie, J. Li, Y. Zhang, *et al.*, "Soliton frequency comb generation in CMOS-compatible silicon nitride microresonators," Photon. Res. **10**(5), 1290–1296 (2022).
27. X. X. Chia and D. T. H. Tan, "Deuterated SiNx: a low-loss, back-end CMOS-compatible platform for nonlinear integrated optics," Nanophotonics **12**(8), 1613–1631 (2023).
28. T. Hiraki, T. Aihara, H. Nishi, *et al.*, "Deuterated SiN/SiON Waveguides on Si Platform and Their Application to C-Band WDM Filters," in IEEE Photonics Journal **9**(5), 1–7 (2017).
29. J. Chiles, N. Nader, D. D. Hickstein, *et al.*, "Deuterated silicon nitride photonic devices for broadband optical frequency comb generation," Opt. Lett. **43**(7), 1527–1530 (2018).
30. D. Bose, M.W. Harrington, A. Isichenko, *et al.*, "Anneal-free ultra-low loss silicon nitride integrated photonics," Light: Sci. Appl. **13**, 156 (2024).
31. A. A. Savchenkov, A. B. Matsko, D. Strekalov, *et al.*, "Low threshold optical oscillations in a whispering gallery mode $CaF_2$ resonator," Phys. Rev. Lett. **93**, 243905 (2004).
32. Ó.B. Helgason, M. Girardi, Z. Ye, *et al.*, "Surpassing the nonlinear conversion efficiency of soliton microcombs," Nat. Photon. **17**, 992–999 (2023).